\documentclass[11pt,letterpaper]{article}
\usepackage[margin=1in]{geometry}
\usepackage[T1]{fontenc}
\usepackage{lmodern}
\usepackage{amsmath,amssymb}
\usepackage{graphicx,booktabs,array}
\usepackage[hidelinks]{hyperref}
\usepackage{microtype}
\usepackage[section]{placeins}
\graphicspath{{figures/}}
\title{Bird-Call Recognition Across Distance:\protect\\Acoustic Degradation and BirdNET Performance}
\author{Hasmik Gharibyan\\\small\texttt{hghariby@usc.edu} \and Bhaskar Krishnamachari\\\small\texttt{bkrishna@usc.edu}}
\date{Viterbi School of Engineering\\University of Southern California}
\begin{document}
\maketitle
\begin{abstract}
As bird calls weaken with distance, their recordings lose spectral detail. We examine how species recognition varies with distance using controlled playback of ten-second excerpts from 20 bird species at seven distances from 1 to 48 ft. An STM32-based recording system acquires and stores the audio, which is subsequently analyzed on a computer using acoustic measurements and BirdNET. Across four metadata configurations, mean-aggregated top-1 accuracy is 95--100\% through 24 ft, 90--95\% at 36 ft, and 80--90\% at 48 ft, with 20 recordings evaluated per configuration at each distance. Across the screened recordings, the median upper spectral edge is below its 1-ft reference at every evaluated distance from 4 to 36 ft, indicating an overall reduction in upper-frequency content with distance. These results show that recognition remains high through 24 ft despite measurable spectral degradation, followed by greater and configuration-dependent losses at longer distances.
\end{abstract}

\section{Introduction}
Automated acoustic monitoring uses recorded vocalizations to identify bird species. BirdNET is a deep-learning system that assigns predicted bird-species labels to recorded sounds~\cite{kahl2021birdnet}. A recording system's usefulness for this task depends on whether those labels remain accurate as the received calls change. Acoustic measurements describe changes in sound level and spectral structure, while recognition accuracy measures how often the expected species is identified.

Distance changes the signal available to the classifier. Calls become weaker, and portions of their spectrum can become less visible against the recording background. We study how well an automated system identifies species as these changes occur. Measuring acoustic characteristics and recognition on the same recordings lets us examine whether visible degradation accompanies a change in task performance.

We conducted a controlled playback study using preprocessed, standardized excerpts from 20 bird species at seven distances from 1 to 48 ft, using each excerpt's nearest recording as its acoustic reference. An STM32-based recording system recorded the playbacks of bird vocalizations for acoustic analysis and BirdNET inference on a computer. Across the screened recordings, the median upper spectral edge was below its 1-ft reference at every evaluated distance from 4 to 36 ft. Mean-aggregated BirdNET predictions identify 19 or 20 of the 20 excerpts at each distance through 24 ft. At 36 ft, BirdNET identifies 18 excerpts without location metadata and 19 with location metadata; at 48 ft, it identifies 16 and 18 excerpts, respectively. The annotated call intervals define the portions used for recognition scoring.

Previous playback studies have evaluated BirdNET recognition across distances, species, recorder types, and environmental conditions~\cite{perez2023distance,recorder2024}. We examine how increasing recording distance affects different bird vocalizations under relatively consistent recording conditions by capturing signals with an STM32-based embedded system at distances from 1 to 48 ft and evaluating how the resulting acoustic degradation influences downstream BirdNET recognition performance.

The contributions of this work are:
\begin{enumerate}
\item A controlled playback experiment spanning excerpts from 20 species and seven distances, with acoustic analysis and BirdNET evaluation applied to the same recorded segments.
\item An empirical comparison of spectral change and species recognition across distance. Representative upper-frequency structures weaken while mean-aggregated top-1 accuracy remains at 95--100\% through 24 ft. Per-configuration counts show the decline at 36 and 48 ft and the effect of aggregation and supplied metadata.
\item An STM32 acquisition workflow with digital active-call level standardization, temporal alignment, and consistent use of active-call annotations for analysis and recognition scoring.
\end{enumerate}

Section~\ref{sec:related} places the study in context. Section~\ref{sec:system} describes the recorder and stimulus preparation. Section~\ref{sec:experiment} presents the experimental protocol and analysis. Section~\ref{sec:results} presents the results, and Section~\ref{sec:discussion} discusses their interpretation and extensions.

\section{Related Work}\label{sec:related}

\subsection{Acoustic Propagation and Distance-Dependent Degradation}

Bird vocalizations undergo acoustic degradation as they propagate through the environment, with attenuation affecting signal amplitude, frequency content, and temporal structure. The extent of this degradation depends not only on propagation distance but also on the acoustic characteristics of the vocalization and the surrounding environment. Priyadarshani et al.~\cite{priyadarshani2018environment} investigated these effects through controlled playback experiments involving 20 bird calls from 11 species under different habitat, transmission-height, and weather conditions. Their findings demonstrated that attenuation is frequency-dependent and influenced by environmental conditions, with higher-frequency components particularly susceptible to attenuation and masking in forest environments. These results establish that increasing distance can alter the spectral characteristics of recorded vocalizations rather than simply reducing their overall amplitude.

The implications of acoustic degradation for passive monitoring have also been examined in studies of species-specific detection ranges. Winiarska et al.~\cite{winiarska2024detection} evaluated vocalizations from 31 European forest bird species along two 500-m transects, investigating how habitat, source height, and seasonal vegetation changes influence detection probability and maximum detection distance. Their results demonstrated substantial variation across species and recording conditions, emphasizing that acoustic detection range cannot be represented by a single distance applicable to all vocalizations. While these studies characterize how bird vocalizations change during propagation and how far they can be detected, an additional question is how such changes affect automated species recognition.

\subsection{Automated Bird Recognition and Distance-Dependent Performance}

Advances in deep learning have enabled automated identification of bird vocalizations from large acoustic datasets. Kahl et al.~\cite{kahl2021birdnet} introduced BirdNET, a deep neural network for avian sound classification, and evaluated its performance using single-species recordings and environmental soundscapes. Their results demonstrated the potential of deep-learning approaches for automated biodiversity monitoring while highlighting the challenges posed by background noise and overlapping vocalizations.

Controlled playback experiments have subsequently investigated the effects of distance and recording conditions on BirdNET recognition. P\'erez-Granados~\cite{perez2023distance} evaluated three bird species using vocalizations broadcast at distances ranging from 10 to 150 m. Recognition performance generally declined with increasing distance and differed among species, with mean recall decreasing from 92.2\% at distances of 50 m or less to 34.9\% beyond 50 m. Although the study found no significant differences between the recorder types tested, its findings indicate that BirdNET recognition performance varies with propagation distance and species characteristics. Somervuo et al.~\cite{somervuo2023effects} investigated the effects of landscape and recording distance on automated bird-species classification using measured acoustic impulse responses to simulate sound propagation. Their work examined how incorporating propagation effects into classification data influences recognition performance, providing a complementary approach to studying distance-dependent acoustic degradation and automated identification.

\subsection{Positioning of the Present Study}

Previous research has demonstrated that bird vocalizations undergo distance-dependent acoustic degradation and that automated recognition performance varies across species and recording conditions. Building on these findings, the present study evaluates 20 selected bird-call excerpts representing 20 species with varying initial BirdNET identification confidence scores across seven recording distances from 1 to 48 ft. This experimental design allows recognition performance to be examined across vocalizations with different baseline recognition characteristics under relatively consistent recording conditions. In addition, the study applies acoustic measurements and BirdNET recognition to the same recordings, enabling a direct comparison between changes in spectral characteristics and identification accuracy as distance increases. The experiment uses an STM32-based recording system integrating an I2S MEMS microphone and microSD storage for audio acquisition. By combining multi-species playback, embedded audio acquisition, and acoustic and classification analyses, this work characterizes the extent to which BirdNET recognition remains accurate despite measurable spectral degradation under the tested conditions.

\section{Recorder and Stimulus Preparation}\label{sec:system}
\subsection{Recording system}
We built the recorder around a NUCLEO-L476RG board, an Adafruit I2S MEMS microphone, a microSD card, and an Adafruit microSD breakout board. The microphone supplies digital audio through I2S. The microcontroller writes audio to the card through SPI using the FatFs filesystem layer. The STM32L476RGT6 has 1 MB of flash memory and 128 KB of SRAM. External storage accommodates recordings that extend beyond the capacity of the temporary audio buffers. Figure~\ref{fig:wiring} in Appendix~\ref{app:preparation} shows the recorder connections.

The final recorded files use one channel, 16-bit PCM samples, and a sampling rate of 35,714 Hz. At this rate, the Nyquist frequency is approximately 17.857 kHz. We use 15 kHz as the practical upper analysis boundary specified for the experiment. The playback files use 44,100 Hz.

We checked the recording frequency scale using tones played by a phone at a distance of 1 inch. The six nominal frequencies were 4, 6, 8, 10, 12, and 14 kHz. For each recorded tone, the dominant detected frequency was compared with the nominal frequency. This test showed detected frequencies of 3,999.6, 5,998.4, 7,994.8, 9,996.6, 11,995.6, and 13,995.4 Hz for nominal inputs of 4 to 14 kHz. Absolute differences range from 0.4 to 5.2 Hz, with the largest difference at 8 kHz, approximately 0.065\% of that nominal frequency. These measurements support the use of the recorder's frequency axis over the tested tones.
%

\subsection{Source recordings and active intervals}
We selected 20 species using a Los Angeles species list and the availability of suitable recordings in the Macaulay Library. The selection favored recordings with clear target vocalizations and relatively low background sound. Appendix~\ref{app:sources} lists the species and media identifiers. For each source recording, a ten-second interval was manually selected to contain clear target vocalizations and, where possible, maximize the amount of target vocal activity within the fixed duration. The selected interval was then extracted using a Python script. Within each ten-second segment, active-call intervals were manually annotated by jointly inspecting the spectrogram and listening to the audio. The annotation boundaries were subsequently reviewed and refined to align with the observed onset and offset of the target vocalizations. These intervals provide a common temporal reference for stimulus preparation, level standardization, acoustic analysis, and recognition evaluation.

\subsection{Playback preparation}
Each selected source recording was processed to produce a standardized ten-second playback stimulus. The source audio was filtered using reviewed frequency cutoffs, converted to mono, resampled to 44.1 kHz, and saved as PCM16 WAV. Samples outside the manually annotated active-call intervals were set to zero, with short fades applied at the interval boundaries to reduce abrupt transitions. Appendix~\ref{app:preparation} gives the filter design, channel-selection rules, resampling and fade parameters, and examples.

To standardize the digital level of the target vocalizations, RMS amplitude was calculated using only samples within the annotated active-call intervals. Let $x[n]$ denote an audio sequence and let $\mathcal{A}$ be the set of sample indices in the annotated active intervals. For samples represented relative to digital full scale, the active-call root-mean-square (RMS) amplitude is
\begin{equation}
 R_{\mathcal A}=\sqrt{\frac{1}{|\mathcal A|}\sum_{n\in\mathcal A}x[n]^2}.
\end{equation}
The corresponding level is $20\log_{10}R_{\mathcal A}$ dBFS. Using only the annotated active samples prevents differences in silent duration from affecting the level estimate.

Each playback file was then scaled to an active-call RMS target of $-25$ dBFS. With the target amplitude $R_{\mathrm{target}}=10^{-25/20}$, the applied gain was
\begin{equation}
 g=\frac{R_{\mathrm{target}}}{R_{\mathcal A}},\qquad y[n]=g x[n].
\end{equation}
A single gain scales the complete recording and retains the relative amplitude pattern within the call. The permitted full-file peak was $-1$ dBFS. The normalization procedure accepted files whose required gain satisfied that peak limit and reported an error otherwise.

The preparation checks cover format, duration, active-call RMS, peak level, and the maximum amplitude in inactive intervals. Appendix~\ref{app:validation} collects the numerical criteria. The checks define the intended stimulus properties and provide separate reports for the processing stages.

As a final check of the prepared playback stimuli, each ten-second source excerpt was evaluated using the BirdNET demo before the distance experiment. This baseline analysis characterized the recognizability of the selected and processed vocalizations after filtering, resampling, silencing, and level normalization. The demo confidence is used only to characterize the source-excerpt baseline and is distinct from the BirdNET-Analyzer scores calculated later from the STM32 recordings.
\begin{table}[!htbp]
\centering
\small
\caption{BirdNET Demo confidence for the normalized source bird-call recordings prior to the distance experiment.}
\begin{tabular}{lr@{\qquad}lr}
\toprule
\textbf{Bird species} & \textbf{Conf. (\%)} &
\textbf{Bird species} & \textbf{Conf. (\%)} \\
\midrule
American Robin            & 94.0 & House Sparrow             & 62.6 \\
Anna's Hummingbird        & 99.8 & Lesser Goldfinch          & 39.8 \\
Barn Swallow              & 94.2 & Mourning Dove             & 86.1 \\
Bewick's Wren             & 62.2 & Northern House Wren       & 86.0 \\
Black Phoebe              & 99.9 & Northern Yellow Warbler   & 70.6 \\
California Towhee         & 86.1 & Northern Mockingbird      & 23.2 \\
Dark-eyed Junco           & 73.2 & Oak Titmouse              & 94.4 \\
European Starling         & 60.9 & Rock Pigeon               & 93.6 \\
Hooded Oriole             & 97.5 & Song Sparrow              & 70.1 \\
House Finch               & 64.6 & Spotted Towhee            & 83.0 \\
\bottomrule
\end{tabular}
\end{table}
\section{Experimental Design and Analysis}\label{sec:experiment}
\subsection{Playback and acquisition}
We conducted the field recordings in Glendale, California, on August 2, 2026, between approximately 7:00 and 9:00 a.m. Conditions were relatively quiet, but we did not independently document an absence of natural bird activity. Natural vocalizations resembling the playback species may therefore have been present. In particular, House Finches, Anna's Hummingbirds, and Mourning Doves may have been vocalizing independently or may have been stimulated by the playback, but no behavioral observations were collected to verify either possibility. 

We used a JBL Boombox speaker with its playback setting fixed at 50 out of 100. The same source sequence and speaker setting were used at seven speaker-to-microphone distances: 1, 4, 8, 12, 24, 36, and 48 ft. We measured A-weighted sound pressure level at 1 ft during the active portions of each excerpt using a Goldeep Digital SI750 sound level meter in Fast mode (0.125 s time response), with a stated accuracy of $\pm1.5$ dB, following the manufacturer's instructions. The reported mean levels across the 20 excerpts ranged from 62.6 to 70.3 dBA, with an average of 66.8 dBA; the excerpts therefore did not have identical physical playback levels. These measurements describe acoustic sound pressure measured in the field.
\begin{table}[htbp]
  \centering
  \caption{A-weighted sound-pressure measurements for the active-call portions of the playback excerpts at 1~ft. The number of readings varied among excerpts. Mean and maximum values are reported directly from the measurement dataset.}
  \label{tab:birdsong-dba-at-1ft}
  \begin{tabular}{lrrr}
    \toprule
    Species & Readings ($n$) & Mean (dBA) & Maximum (dBA) \\
    \midrule
    American Robin          & 10 & 66.750 & 77.50 \\
    Anna's Hummingbird      & 12 & 68.683 & 73.80 \\
    Barn Swallow            & 13 & 69.954 & 75.10 \\
    Bewick's Wren           &  4 & 65.800 & 74.50 \\
    Black Phoebe            &  6 & 67.250 & 74.10 \\
    California Towhee       &  3 & 65.967 & 70.30 \\
    Dark-eyed Junco         &  3 & 66.600 & 70.40 \\
    European Starling       & 13 & 65.238 & 74.30 \\
    Hooded Oriole           &  8 & 62.613 & 74.90 \\
    House Finch             &  9 & 67.511 & 78.70 \\
    House Sparrow           & 12 & 63.925 & 71.00 \\
    Lesser Goldfinch        & 13 & 68.377 & 77.30 \\
    Mourning Dove           & 10 & 68.940 & 80.30 \\
    Northern House Wren     &  9 & 68.511 & 73.80 \\
    Northern Yellow Warbler & 10 & 66.560 & 74.20 \\
    Northern Mockingbird    &  2 & 66.650 & 75.40 \\
    Oak Titmouse            &  6 & 70.317 & 74.30 \\
    Rock Pigeon             &  6 & 67.650 & 78.60 \\
    Song Sparrow            &  4 & 66.375 & 74.30 \\
    Spotted Towhee          &  5 & 62.660 & 71.00 \\
    \bottomrule
  \end{tabular}
\end{table}

The evaluated dataset contains 20 species segments at each distance, giving 140 ten-second recordings and 560 recording--configuration evaluations. Every evaluation has status OK and enters the recognition summaries. The analysis artifacts are archived at source commit \texttt{5489ac6}~\cite{gharibyan2026artifacts}. The source sequence includes one selected excerpt per species, and the evaluated set contains one segment per species and distance.

We combined the prepared stimuli into a fixed playback sequence. It contained five seconds of initial silence, ten seconds for each of 20 birds, five seconds between successive birds, and five seconds of final silence. The total duration was $5+20(10)+19(5)+5=305$ seconds. Before combination, each input file was checked for one channel, a 44,100 Hz sample rate, 441,000 samples, finite sample values, and amplitudes within digital full scale.

For each distance, the recorder captured the combined sequence as one continuous WAV file on the SD card. We transferred these files to a computer for alignment, segmentation, acoustic analysis, and BirdNET inference. The STM32 performs acquisition and storage in this workflow. BirdNET processing takes place on the computer.

We combined the prepared stimuli into a fixed playback sequence. It contained five seconds of initial silence, ten seconds for each of 20 birds, five seconds between successive birds, and five seconds of final silence. The total duration was $5+20(10)+19(5)+5=305$ seconds. Before combination, each input file was checked for one channel, a 44,100 Hz sample rate, 441,000 samples, finite sample values, and amplitudes within digital full scale.

For each distance, the recorder captured the combined sequence as one continuous WAV file on the SD card. We transferred these files to a computer for alignment, segmentation, acoustic analysis, and BirdNET inference. The STM32 performs acquisition and storage in this workflow. BirdNET processing takes place on the computer.

\subsection{Temporal alignment and segmentation}
The continuous recordings were aligned to the known playback sequence to locate the ten-second species segments. Global correlation first estimated the playback start within the recording. The playback schedule then supplied an expected start for each bird. For bird index $i\in\{1,\ldots,20\}$, the scheduled start relative to the playback origin is
\begin{equation}
 t_i=5+15(i-1)\ \mathrm{s}.
\end{equation}
Adding the estimated global offset gives the predicted position in the recorded file. Local correlation refined each bird's position within an approximately $\pm2$ s search region. Reference signals were temporarily resampled to the recording time scale for alignment. The output segments were cut from the original recorded samples.

\begin{figure}[tb]
\centering
\includegraphics[width=\linewidth]{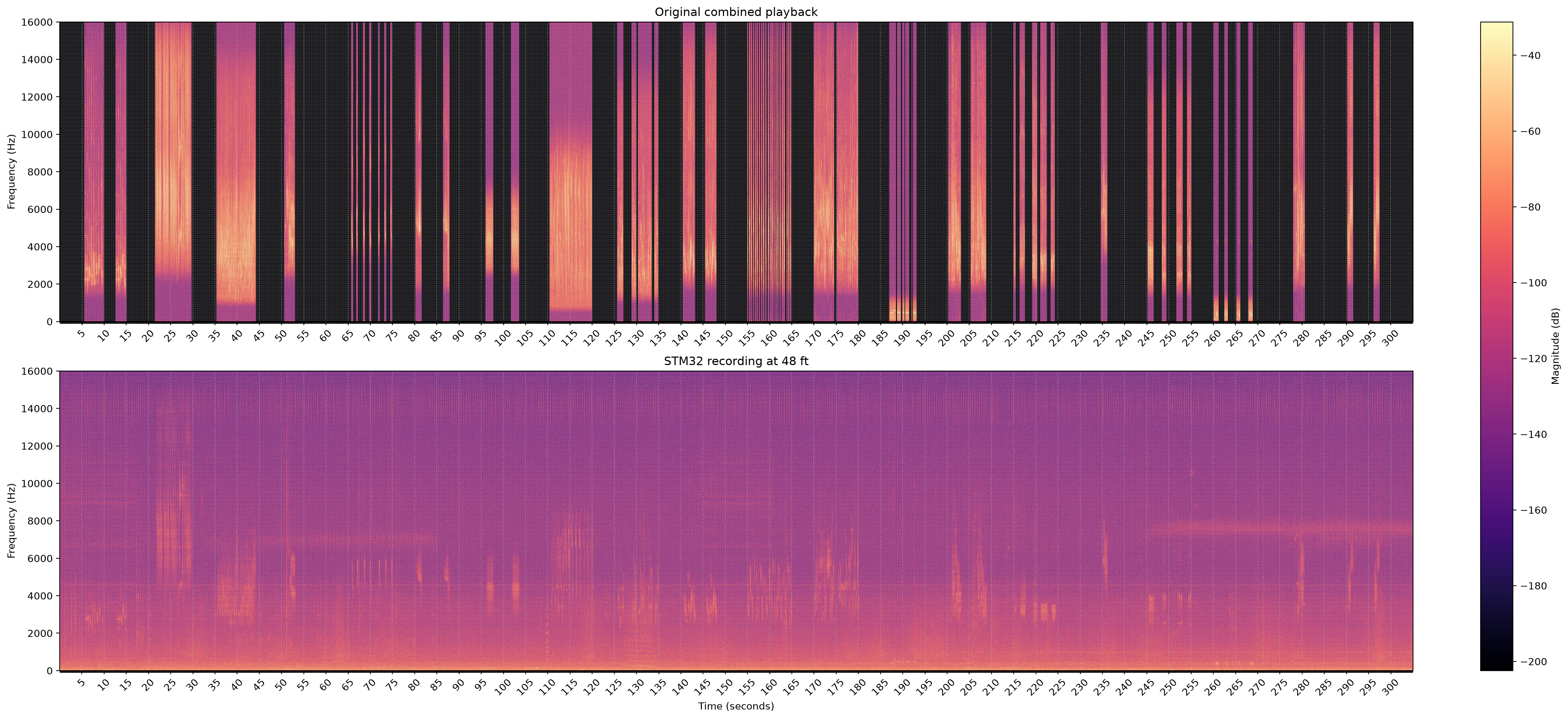}
\caption{Full spectrograms of the original combined playback sequence and the STM32 recording at 48 ft. The playback order supplies the temporal context used to identify species search regions before constrained correlation.}\label{fig:alignment48}
\end{figure}

We used this global-plus-local procedure for the 1--36 ft recordings. For the 48-ft recording, the final procedure used search regions selected through visual comparison of the full playback and recorded spectrograms. Cross-correlation within each selected region then determined the segment position. Figure~\ref{fig:alignment48} shows the full-length comparison used to guide this step. Each extracted segment was expected to contain 357,140 samples, corresponding to ten seconds at 35,714 Hz.

At 48 ft, the lower recorded signal strength made the selection of local correlation peaks sensitive to background events. A tested global timing relationship gave a global correlation of approximately 0.0129 at 48 ft.

\FloatBarrier
\subsection{Recording validation}
We checked the extracted files for sampling rate, duration, channel count, PCM sample width, and frame count. Signal checks included sample extrema, mean, RMS, peak, clipping fraction, near-full-scale fraction, and low-byte diversity. Calibration recordings had absolute mean DC offsets between 6,426 and 6,437 PCM counts. A threshold of 7,000 counts was selected for flagging unusually large offsets. Acoustic measurements were computed after removing the constant DC component.

Recordings with more than 0.1\% of samples at the PCM16 limits or more than 1\% near full scale were classified as failures. The near-full-scale threshold was 32,000 counts in magnitude. Fewer than 32 distinct low-byte values produced a warning for review. Each recording received a PASS, WARNING, or FAIL status. Validation records identify successfully analyzed recordings within the expected segment set.

\subsection{Acoustic measurements}
We analyzed each species within its annotated active intervals and its selected frequency range. The 1-ft recording of that species served as the reference. Welch power spectral density (PSD) estimates used a 0.10-second Hann window with 50\% overlap. We calculated active RMS, peak, total band power, and their changes relative to the reference. The reference PSD also defined low, middle, and high bands with equal integrated energy. These reference boundaries were retained across distances.

The spectral examples use a common magnitude scale within each species to compare visible call structure across distance. The analysis also calculated spectral shape and energy-distribution metrics relative to 1 ft. Appendix~\ref{app:acoustic} gives their definitions and the heuristic screening score used to select recordings for descriptive spectral summaries.

\subsection{BirdNET configurations and scoring}
We processed the recording set under four configured combinations of location and week information, shown in Table~\ref{tab:metadata}. The settings include the Los Angeles coordinates $(34.0522,-118.2437)$ and a week parameter of 27. Week 27 was selected as a representative summer period during which all 20 selected species were expected to occur in the Los Angeles area. A value of $-1$ was used for omitted metadata in the runs. We invoked the Python \texttt{birdnet\_analyzer.analyze} interface with the top 20 predictions, a requested minimum confidence of 0.01, eight threads, and a batch size of one. Saved analysis parameters specify 48,000 Hz model input, 3-second segments, zero segment overlap, a 1-second minimum segment, a 0--15,000 Hz bandpass, audio speed 1.0, and enabled merging of consecutive detections. File splitting was set to 600 seconds, and the default classifier was used.

\begin{table}[tb]
\centering
\caption{Metadata values configured for the four BirdNET runs. The labels describe the supplied inputs.}\label{tab:metadata}
\begin{tabular}{llll}
\toprule
Run & Latitude & Longitude & Week\\
\midrule
E1: no metadata & $-1$ & $-1$ & $-1$\\
E2: week only & $-1$ & $-1$ & 27\\
E3: location only & 34.0522 & $-118.2437$ & $-1$\\
E4: week and location & 34.0522 & $-118.2437$ & 27\\
\bottomrule
\end{tabular}
\end{table}

BirdNET-Analyzer provides geographic and seasonal species filtering~\cite{birdnetdocs}. Table~\ref{tab:metadata} records the inputs supplied to that process. The saved detections include confidences below 0.01 under the top-20 configuration.

We selected BirdNET windows that overlapped a known active-call interval by at least 0.25 s. For these windows, the evaluation recorded the expected species' confidence and rank and checked its presence in the top 1, top 3, and top 5 predictions. Name mappings matched Northern House Wren to House Wren and Northern Yellow Warbler to Yellow Warbler in the BirdNET output.

Recording-level predictions were obtained by aggregating window scores. Let $\mathcal W_r$ be the active windows for recording $r$. For mean aggregation, let $c_{s,w}^{(5)}$ be the confidence of species $s$ when it appears among the top five detections of window $w$, with zero contribution otherwise. The recording score is
\begin{equation}
 C^{\mathrm{mean}}_{s,r}=\frac{1}{|\mathcal W_r|}\sum_{w\in\mathcal W_r}c_{s,w}^{(5)}.
\end{equation}
This score combines confidence and consistency across the active windows. Maximum aggregation uses a species' highest confidence among all retained detections in eligible active windows. Equal recording scores are ordered alphabetically by species name. Ranking the species by either score produces recording-level top-$k$ predictions. These two scoring procedures differ in both the aggregation rule and the candidate pool: mean aggregation uses each window's top five detections, while maximum aggregation uses all retained detections, up to 20 per window.

The analysis also calculated active-window detection rate, expected-species confidence summaries, and confidence change and retention relative to 1 ft. Retention is the ratio between a recording's expected-species mean confidence and the corresponding 1-ft confidence. Its interpretation requires a positive reference confidence. Accuracy uses the number of correct eligible recordings divided by the number evaluated within each species, distance, and configuration group. Windows enter the evaluation through the saved detection rows. A recording with zero eligible active windows receives an ERROR status. All 140 recordings have eligible windows under every configuration. The recognition plots used rows with status = OK. Each overall distance summary pooled results across birds and the four metadata configurations. These repeated evaluations share the underlying audio recordings.

\section{Results}\label{sec:results}
\subsection{BirdNET recognition across distance}
The BirdNET web-demo results provided a pre-playback baseline for the normalized source excerpts. Expected-species confidence ranged from 23.2\% for Northern Mockingbird to 99.9\% for Black Phoebe, with a mean of 77.1\% and a median of 84.5\% across the 20 excerpts. Eighteen excerpts had expected-species confidence of at least 50\%, including seven at or above 90\%. The normalized ten-second excerpts were submitted without location or seasonal metadata and included the inactive intervals between annotated vocalizations. These values are therefore used only to characterize baseline source-excerpt recognizability and are not treated as directly equivalent to the BirdNET-Analyzer scores obtained from the STM32 recordings. A species-level comparison between the source-demo baseline and the later BirdNET-Analyzer confidence trajectories is provided in Appendix~E.

Table~\ref{tab:recognition} gives recording-level top-1 accuracy for both aggregation methods under all four metadata configurations. Each configuration evaluates the same 20 recordings at each distance. Mean aggregation achieves 95--100\% accuracy from 1 through 24~ft, 90--95\% at 36~ft, and 80--90\% at 48~ft. Maximum aggregation achieves 95--100\% through 24~ft, 85--90\% at 36~ft, and 75--90\% at 48~ft.
\begin{table}[htbp]
\centering\small
\caption{Recording-level top-1 accuracy by distance and metadata configuration. Each cell gives correct/evaluated recordings (percent). E1--E4 use the inputs in Table~\ref{tab:metadata}. All configurations score the same 20 recordings at each distance.}\label{tab:recognition}
\begin{tabular}{rcccc}
\toprule
\multicolumn{5}{c}{Mean aggregation}\\
Distance (ft) & E1 & E2 & E3 & E4\\
\midrule
1 & 19/20 (95\%) & 19/20 (95\%) & 19/20 (95\%) & 19/20 (95\%)\\
4 & 20/20 (100\%) & 20/20 (100\%) & 20/20 (100\%) & 20/20 (100\%)\\
8 & 19/20 (95\%) & 19/20 (95\%) & 20/20 (100\%) & 20/20 (100\%)\\
12 & 20/20 (100\%) & 20/20 (100\%) & 20/20 (100\%) & 20/20 (100\%)\\
24 & 19/20 (95\%) & 19/20 (95\%) & 20/20 (100\%) & 20/20 (100\%)\\
36 & 18/20 (90\%) & 18/20 (90\%) & 19/20 (95\%) & 19/20 (95\%)\\
48 & 16/20 (80\%) & 16/20 (80\%) & 18/20 (90\%) & 18/20 (90\%)\\
\bottomrule
\end{tabular}
\par\medskip
\begin{tabular}{rcccc}
\toprule
\multicolumn{5}{c}{Max aggregation}\\
Distance (ft) & E1 & E2 & E3 & E4\\
\midrule
1 & 19/20 (95\%) & 19/20 (95\%) & 19/20 (95\%) & 19/20 (95\%)\\
4 & 19/20 (95\%) & 19/20 (95\%) & 19/20 (95\%) & 19/20 (95\%)\\
8 & 19/20 (95\%) & 19/20 (95\%) & 20/20 (100\%) & 20/20 (100\%)\\
12 & 19/20 (95\%) & 19/20 (95\%) & 20/20 (100\%) & 20/20 (100\%)\\
24 & 19/20 (95\%) & 19/20 (95\%) & 20/20 (100\%) & 20/20 (100\%)\\
36 & 17/20 (85\%) & 17/20 (85\%) & 18/20 (90\%) & 18/20 (90\%)\\
48 & 15/20 (75\%) & 15/20 (75\%) & 18/20 (90\%) & 18/20 (90\%)\\
\bottomrule
\end{tabular}
\par\medskip
\end{table}

At 48~ft, supplying location metadata increased the mean-aggregation count from 16/20 to 18/20 and the maximum-aggregation count from 15/20 to 18/20. E1 and E2 produced identical recording-level top-1 predictions, as did E3 and E4. Thus, adding week~27 did not change top-1 recognition in this experiment, whereas location information improved the far-distance results.

\begin{figure}[htbp]
\centering
\includegraphics[width=\textwidth]{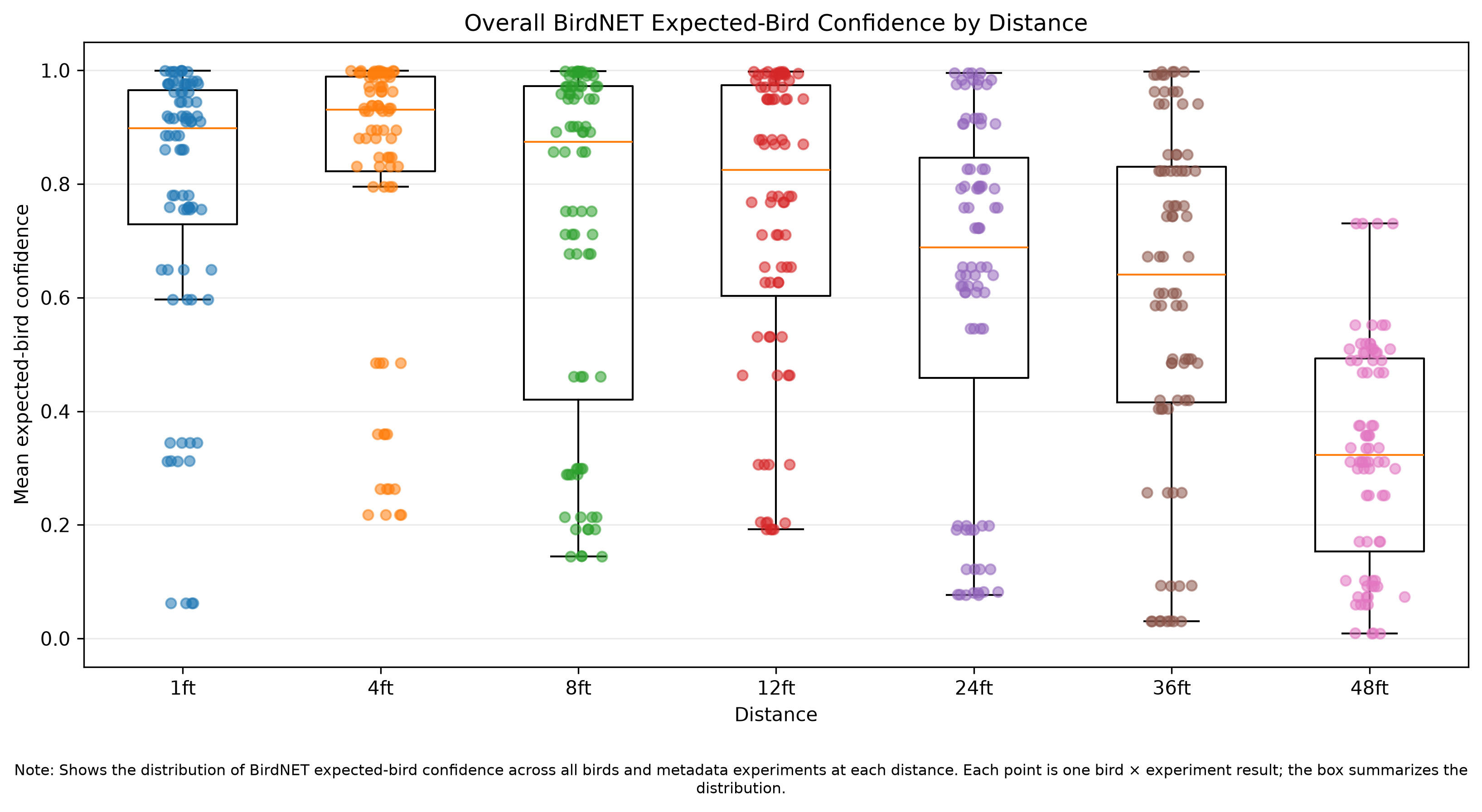}
\caption{Distribution of recording-level expected-species mean confidence by distance. Each point represents one recording--configuration evaluation; each distance contains 80 evaluations derived from 20 recordings processed under four metadata configurations. Boxes summarize the pooled distributions. The repeated configuration evaluations share the same underlying audio recordings.}\label{fig:confidence-distribution}
\end{figure}

The expected-species mean confidence decreased with distance, from approximately 0.780 at 1 ft to 0.326 at 48 ft. Figure~\ref{fig:confidence-distribution} shows the pooled distribution across the four metadata configurations. Although confidence generally declined with distance, the expected species often retained the highest recording-level rank, consistent with the relatively high top-1 accuracy observed through 24 ft.

The pooled distributions conceal substantial variation among individual excerpts. Appendix E therefore presents each metadata configuration separately, with species shown consistently across distances. These plots allow species-excerpt confidence trajectories to be examined without pooling repeated evaluations of the same recordings and show that some excerpts retained comparatively high confidence at longer distances while others declined earlier.

For Lesser Goldfinch, I would not put the American Goldfinch 44.4\% detail in the main paragraph unless you want one short illustrative sentence. It fits better in Appendix E with the species-level discussion.
\FloatBarrier
\subsection{Spectral changes across distance}
\begin{figure}[htbp]
\centering
\includegraphics[width=\textwidth]{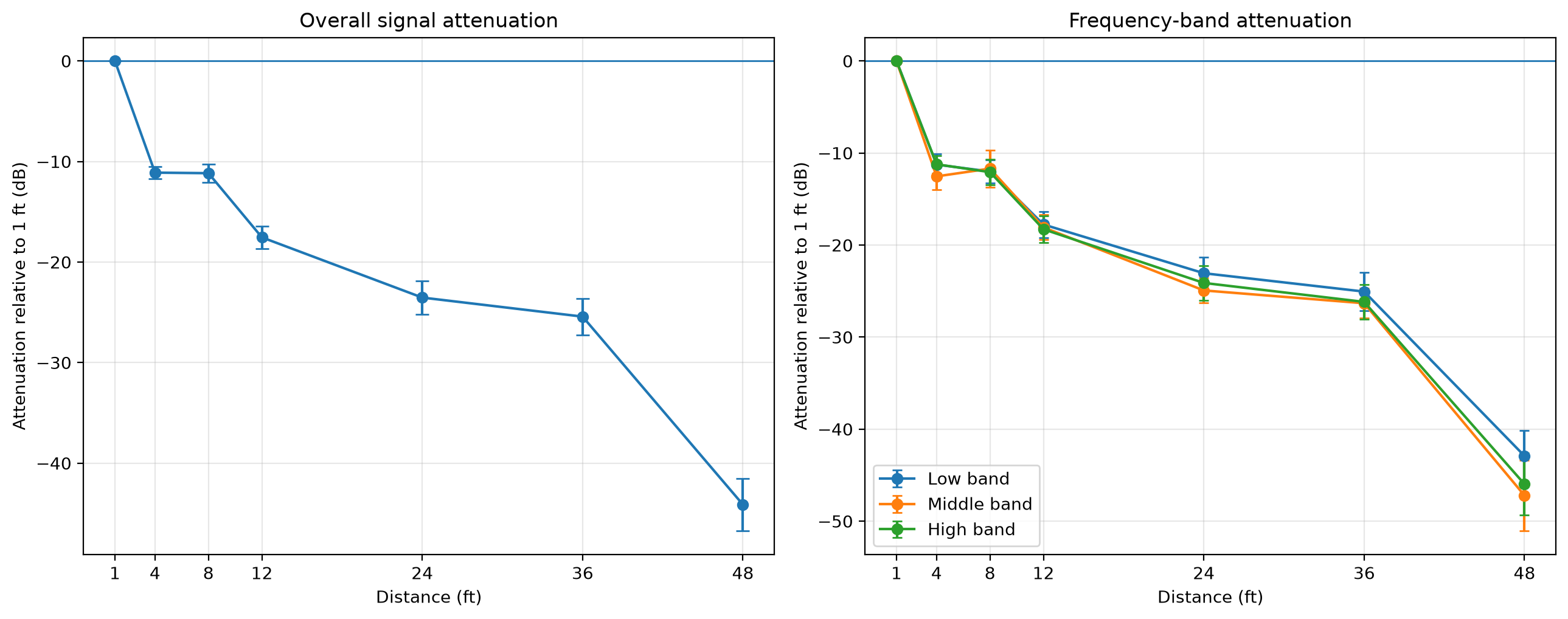}
\caption{Recorded attenuation relative to each excerpt's 1-ft STM32 recording. The left panel shows mean overall attenuation across the 20 excerpts, and the right panel shows mean attenuation in the low-, middle-, and high-frequency bands. Error bars indicate 95\% confidence intervals across excerpts.}\label{fig:total_attenuation}
\end{figure}

Across the 20 excerpts, total spectral power decreased relative to each excerpt's 1-ft reference as recording distance increased. Mean total spectral-power change was approximately $-11.1$~dB at 4~ft, $-11.2$~dB at 8~ft, $-17.5$~dB at 12~ft, $-23.5$~dB at 24~ft, $-25.4$~dB at 36~ft, and $-44.1$~dB at 48~ft. Figure~\ref{fig:total_attenuation} summarizes this trend across all 20 excerpts. Error bars indicate 95\% confidence intervals around the mean across excerpts at each distance.

\begin{figure}[htbp]
\centering
\includegraphics[width=\linewidth,height=.76\textheight,keepaspectratio]{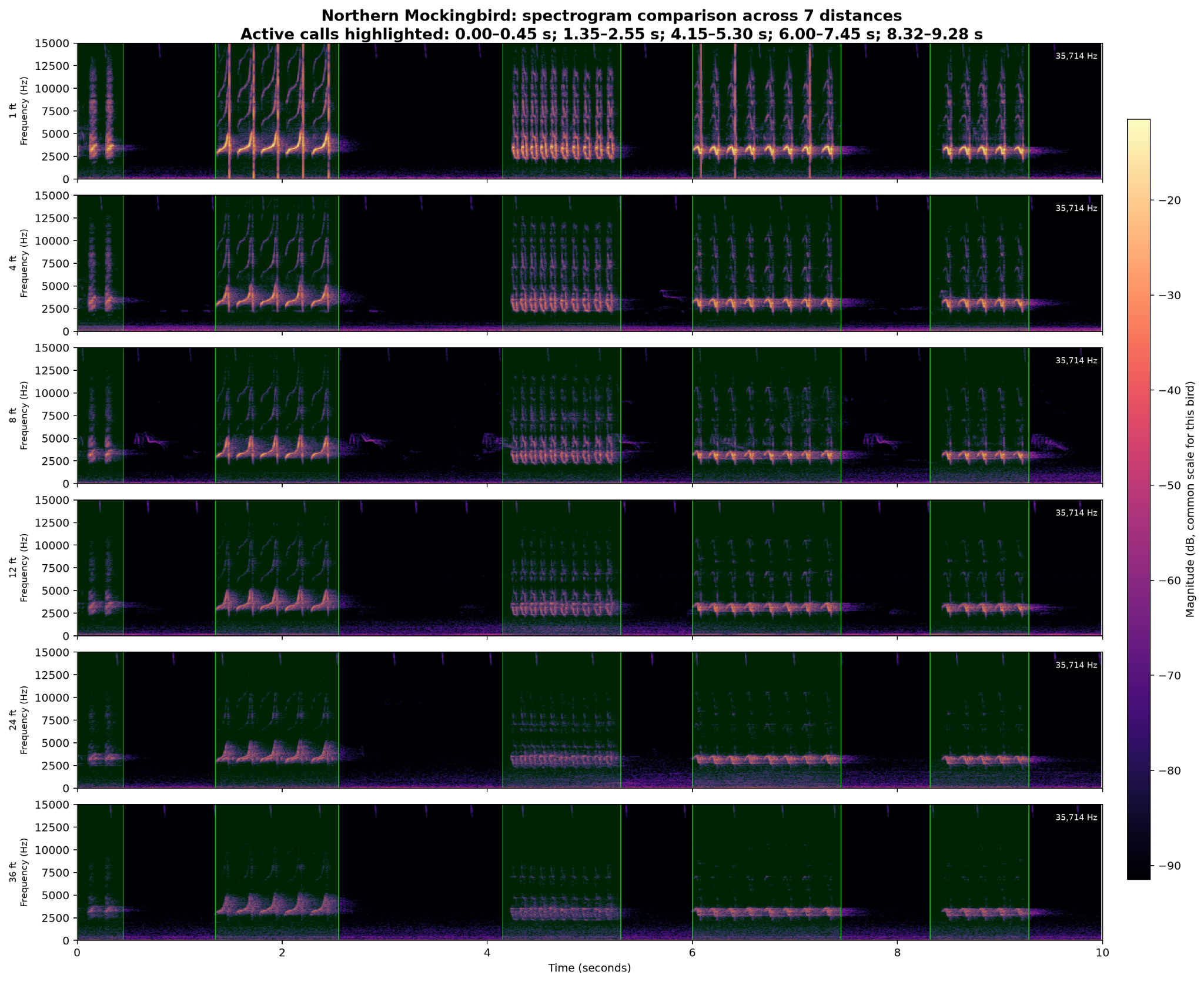}
\caption{Northern Mockingbird recorded at 1, 4, 8, 12, 24, and 36 ft. Green intervals indicate the annotated active calls. The six panels share a magnitude scale and show upper-frequency components becoming less prominent with distance.}\label{fig:mockingbird}
\end{figure}

\begin{figure}[htbp]
\centering
\includegraphics[width=\linewidth,height=.76\textheight,keepaspectratio]{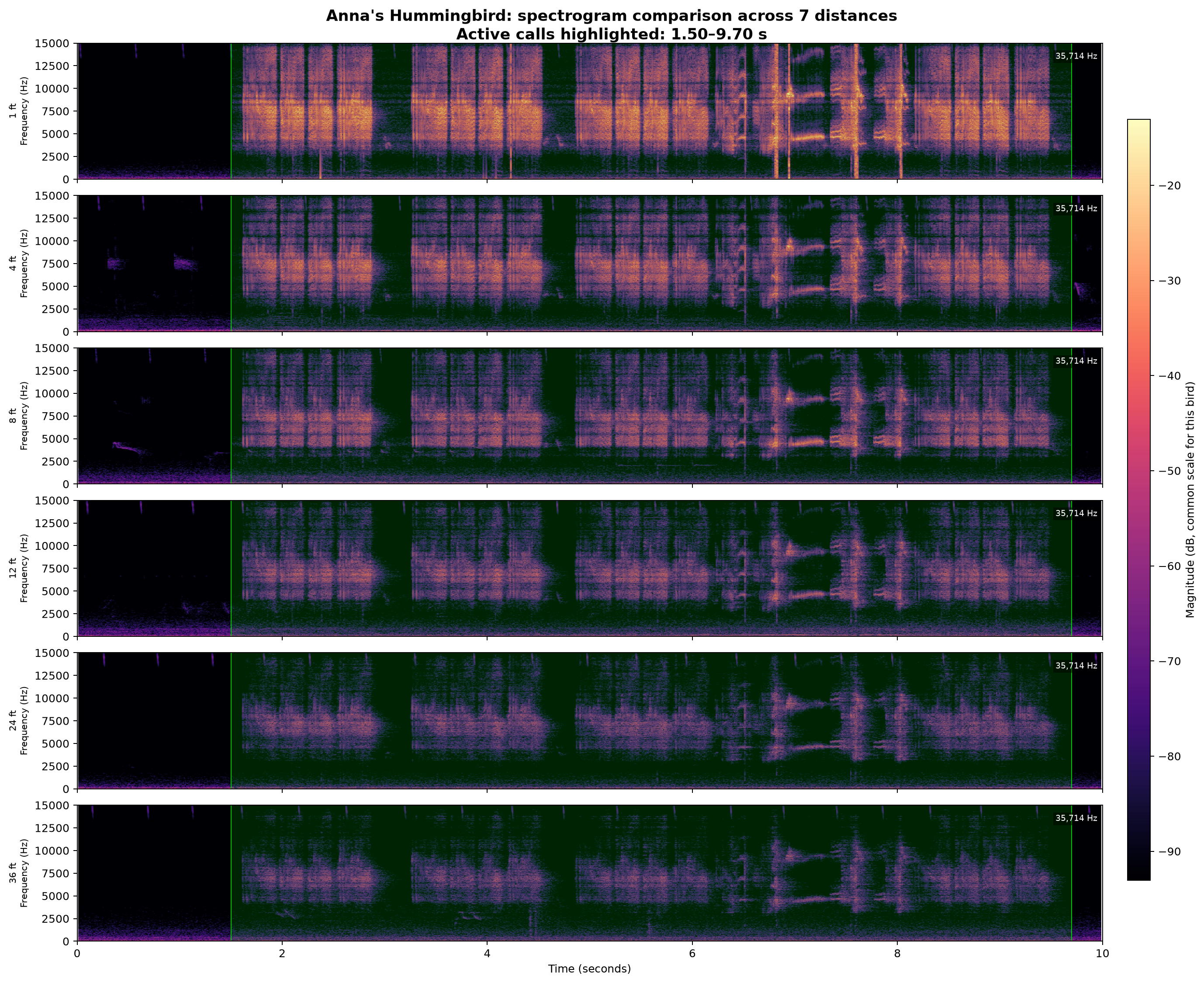}
\caption{Anna's Hummingbird recorded at the six displayed distances of 1, 4, 8, 12, 24, and 36 ft. Green intervals indicate the annotated active calls. The six panels share a magnitude scale and show upper-frequency content becoming less prominent with distance.}\label{fig:hummingbird}
\end{figure}

\FloatBarrier
As an illustration of the spectral changes observed with distance, Figures~\ref{fig:mockingbird} and~\ref{fig:hummingbird} show Northern Mockingbird and Anna's Hummingbird at the six displayed distances from 1 through 36 ft. Within each figure, the panels share a magnitude scale for that species. Call timing remains identifiable across these panels, and upper-frequency components become less prominent as distance increases. In the Northern Mockingbird example, the upper traces are clearer at 1 ft and become progressively weaker in the more distant panels. Anna's Hummingbird likewise retains its visible temporal call pattern while upper-frequency content becomes weaker.

\FloatBarrier
\subsection{Heuristic screening and descriptive spectral statistics}
\begin{figure}[htbp]
\centering
\includegraphics[width=.86\linewidth]{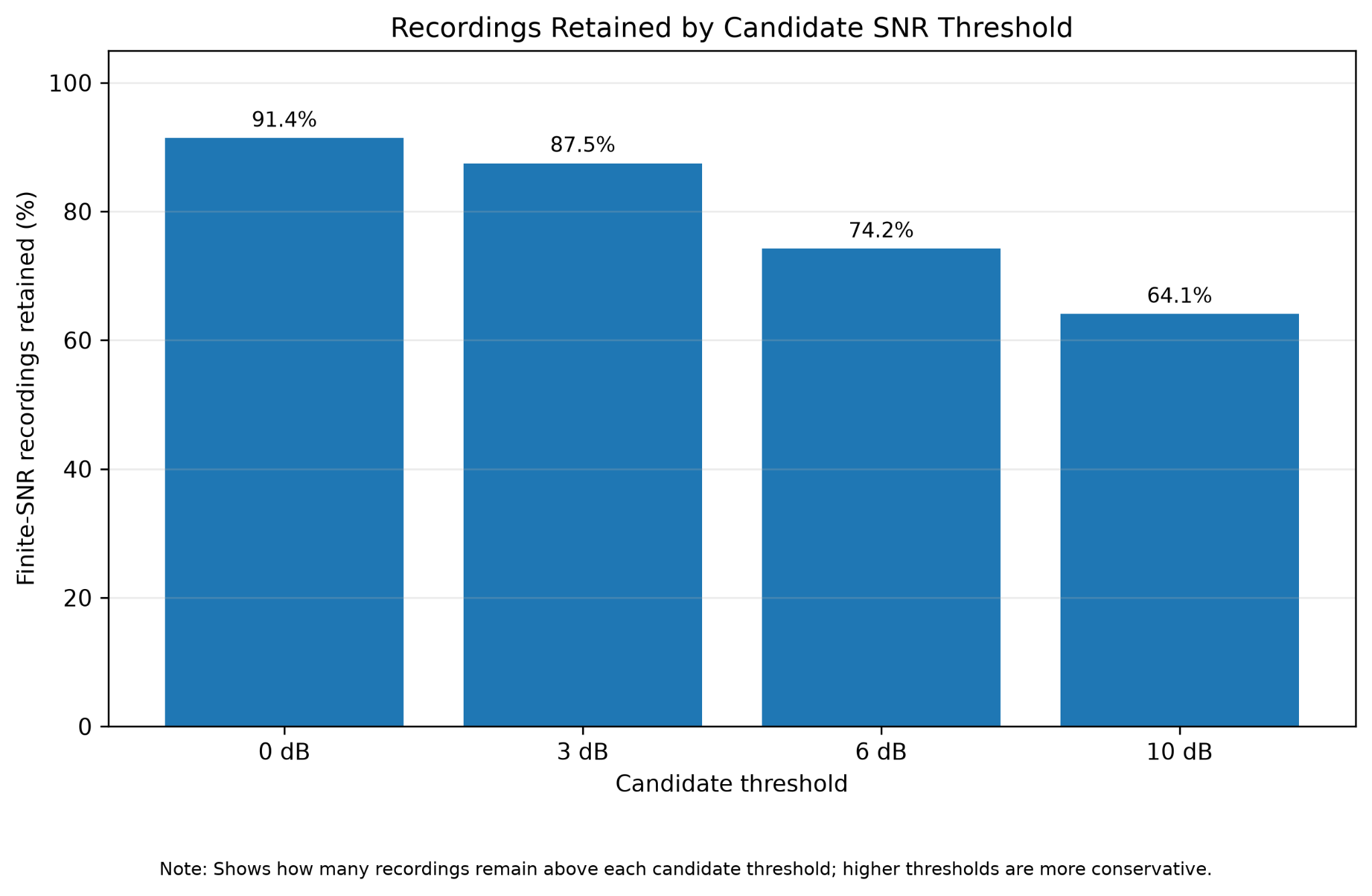}
\caption{Percentage of recordings with finite heuristic screening scores meeting each candidate threshold. The source plot labels the score SNR. The 3 dB criterion selects 87.5\% of that set.}\label{fig:snr}
\end{figure}

We compared thresholds of 0, 3, 6, and 10 dB for the heuristic screening score defined in Appendix~\ref{app:acoustic}. These thresholds select 117, 112, 95, and 82 of the 128 recordings with finite scores, corresponding to 91.4\%, 87.5\%, 74.2\%, and 64.1\%, respectively (Figure~\ref{fig:snr}). We use the 3 dB criterion to define the subset summarized in Table~\ref{tab:acoustic}.

The threshold comparison describes how subset size changes with the screening rule. The Frequency Distortion Index (FDI) measures spectral shape change relative to the 1-ft recording.

\begin{table}[htbp]
\centering\small
\caption{Descriptive spectral statistics for recordings meeting the heuristic screening criterion. Each distance has 20 recordings before selection. FDI and upper spectral-edge changes are medians for the selected subset, using each excerpt\textquotesingle s 1-ft reference. Selection requires a finite score of at least 3 dB. The subset varies with distance.}\label{tab:acoustic}
\begin{tabular}{rrrrr}
\toprule
Distance & Finite score & Score $\geq3$ dB & FDI & Upper-edge change\\
(ft) & (count) & (count) & (dB) & (Hz)\\
\midrule
1 & 20 & 20 & 0.00 & 0\\
4 & 20 & 20 & 5.35 & -109\\
8 & 20 & 20 & 6.71 & -360\\
12 & 20 & 19 & 5.50 & -279\\
24 & 20 & 19 & 4.45 & -197\\
36 & 19 & 13 & 4.59 & -117\\
48 & 9 & 1 & 10.47 & 285\\
\bottomrule
\end{tabular}
\end{table}

Table~\ref{tab:acoustic} presents descriptive statistics for the subset meeting the screening criterion. Of 140 recordings, 128 have finite scores, and 112 meet the 3 dB criterion. The selected counts fall to 13 at 36 ft and one at 48 ft. Within this subset, the median upper spectral edge is lower than its 1-ft reference at every distance from 4 through 36 ft. Median FDI varies between 4.45 and 6.71 dB over those distances. These medians vary non-monotonically, and each distance uses the recordings that meet the criterion at that distance. The 48-ft values describe the single selected recording.

%

%
\section{Discussion}\label{sec:discussion}
\subsection{Interpreting acoustic change and recognition}
Mean-aggregated BirdNET accuracy remains at 95--100\% through 24 ft while the two illustrated excerpts lose visible upper-frequency detail. At 36 and 48 ft, fewer excerpts receive the correct top-1 prediction. The per-configuration results show how recognition varies with distance in this playback experiment. The spectrograms identify changes in the input signal, and the accuracy table measures the species-identification outcome on the same recording set.

At 48 ft, location information improves top-1 accuracy by 10 percentage points for mean aggregation and 15 points for maximum aggregation. Adding the configured week leaves recording-level top-1 predictions unchanged. These comparisons hold the audio fixed and vary the supplied metadata. The benefit of location is therefore specific to these excerpts and configurations.

The study follows selected prepared excerpts through one speaker and recorder across seven distances. This design supports within-excerpt comparisons relative to 1 ft. The known playback schedule and active-call annotations guide segmentation and scoring. At 48 ft, visually selected search regions supply additional timing information for local correlation.

RMS normalization fixes the active-call amplitude of the digital playback files. The acoustic level at the microphone also depends on the speaker's response, the vocalization spectrum, source direction, and propagation conditions. The speaker setting of 50/100 documents an equipment setting. A sound-pressure measurement would provide a physical level reference. The tone test checks frequency correspondence at six frequencies. An amplitude-response measurement would characterize the frequency-dependent gain of the acquisition chain.

The heuristic screening score combines digital mean-square estimates from normalized source audio and recorded audio. Relating the source contribution to acoustic power at the microphone requires the transfer through the speaker, propagation path, microphone, and recording gain. The calculation adds the source term at unity gain and holds it fixed across distances. Its threshold therefore defines a selected subset for descriptive statistics. Table~\ref{tab:acoustic} describes that subset, including a single recording at 48 ft. Recognition accuracy in Table~\ref{tab:recognition} uses all 20 recordings per distance.

\subsection{Extensions}
A future paired playback study could compare the two prepared filtering branches for the 16 selected species. Replaying both branches under the same physical conditions would allow comparison of recognition and spectral measurements. Changing the retained frequency content also changes the gain required to reach the common RMS target, so that comparison would need to account for normalization gain.

Additional source exemplars and independent trials would separate variation among calls, trials, and species. Further experiments could vary background conditions, speaker direction, and recording hardware. Physical sound-level measurements and a measured frequency response would help relate changes in recorded power to the playback and acquisition chain. An independent timing reference would allow measurement of alignment error.

\section{Conclusion}
Across the four metadata configurations, mean-aggregated BirdNET accuracy is 95--100\% through 24 ft, 90--95\% at 36 ft, and 80--90\% at 48 ft. Representative recordings show upper-frequency call components weakening from 1 to 36 ft. Recognition therefore remains accurate for 19 or 20 of the 20 excerpts through 24 ft as visible spectral detail decreases. The STM32 playback experiment places these acoustic observations and species predictions on the same set of recorded excerpts.

\section*{AI Use Statement}
AI tools (specifically ChatGPT/Codex) were used to help edit and format the paper. The human authors accept full responsibility for the contents of this work.
\clearpage
\begingroup
\small
\bibliographystyle{plain}
\bibliography{references}

@article{kahl2021birdnet,
  author = {Kahl, Stefan and Wood, Connor M. and Eibl, Maximilian and Klinck, Holger},
  title = {{BirdNET}: A deep learning solution for avian diversity monitoring},
  journal = {Ecological Informatics},
  volume = {61},
  pages = {101236},
  year = {2021},
  doi = {10.1016/j.ecoinf.2021.101236},
  note = {\url{https://doi.org/10.1016/j.ecoinf.2021.101236}}
}

@article{perez2023distance,
  author = {P{\'e}rez-Granados, Cristian},
  title = {A first assessment of {BirdNET} performance at varying distances: A playback experiment},
  journal = {Ardeola},
  volume = {70},
  number = {2},
  pages = {257--269},
  year = {2023},
  doi = {10.13157/arla.70.2.2023.sc1},
  note = {\url{https://doi.org/10.13157/arla.70.2.2023.sc1}}
}

@article{recorder2024,
  author = {Huang, Wantao and Hao, Zezhou and Zhang, Zixin and Xiao, Zhishu and Zhang, Chengyun},
  title = {A comparison of bird sound recognition performance among acoustic recorders},
  journal = {Biodiversity Science},
  volume = {32},
  number = {10},
  pages = {24273},
  year = {2024},
  note = {\url{https://doi.org/10.17520/biods.2024273}}
}

@article{priyadarshani2018environment,
  author = {Priyadarshani, Nirosha and Castro, Isabel and Marsland, Stephen},
  title = {The impact of environmental factors in birdsong acquisition using automated recorders},
  journal = {Ecology and Evolution},
  volume = {8},
  pages = {5016--5033},
  year = {2018},
  doi = {10.1002/ece3.3889},
  note = {\url{https://doi.org/10.1002/ece3.3889}}
}

@article{winiarska2024detection,
  author = {Winiarska, Dominika and Szyma{\'n}ski, Pawe{\l} and Osiejuk, Tomasz S.},
  title = {Detection ranges of forest bird vocalisations: Guidelines for passive acoustic monitoring},
  journal = {Scientific Reports},
  volume = {14},
  pages = {894},
  year = {2024},
  doi = {10.1038/s41598-024-51297-z},
  note = {\url{https://doi.org/10.1038/s41598-024-51297-z}}
}

@article{somervuo2023effects,
  author = {Somervuo, Panu and Lauha, Petteri and Lokki, Tapio},
  title = {Effects of landscape and distance in automatic audio based bird species identification},
  journal = {The Journal of the Acoustical Society of America},
  volume = {154},
  number = {1},
  pages = {245--254},
  year = {2023},
  doi = {10.1121/10.0020153},
  note = {\url{https://doi.org/10.1121/10.0020153}}
}

@misc{birdnetdocs,
  author = {{BirdNET Team}},
  title = {{BirdNET-Analyzer}: Command line interface},
  howpublished = {\url{https://birdnet-team.github.io/BirdNET-Analyzer/usage/cli.html}},
  note = {Accessed September 5, 2026}
}

@misc{gharibyan2026artifacts,
 author = {Gharibyan, Hasmik},
 title = {{STM32-Audio-Frequency-Analysis}: Experimental artifacts},
 year = {2026},
 howpublished = {\url{https://github.com/hghariby/STM32-Audio-Frequency-Analysis/tree/5489ac6}},
 note = {Commit 5489ac6, accessed September 5, 2026}
}
\endgroup
\clearpage
\appendix
\section{Recorder Connections and Preparation Details}\label{app:preparation}
\begin{figure}[htbp]
\centering
\includegraphics[width=.64\linewidth]{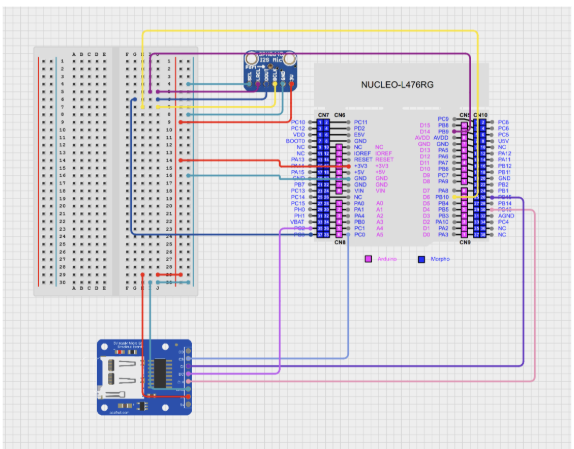}
\caption{Connections among the NUCLEO-L476RG, I2S microphone, and microSD breakout board used for audio acquisition and storage.}\label{fig:wiring}
\end{figure}

\FloatBarrier
\subsection{Frequency selection and filtering}
We estimated the power spectral density (PSD) of the active-call samples using Welch's method with a 0.10-second Hann window and 50\% overlap. The initial frequency search extended from 100 Hz to the source file's Nyquist frequency. Bins more than 40 dB below the peak PSD were excluded from the frequency-range estimate. Within the retained spectrum, the 1st and 99th cumulative-energy percentiles defined an initial interval containing the central 98\% of the retained energy. A 500 Hz margin on either side provided suggested filter boundaries for visual review.

We reviewed these boundaries against the source spectrograms and selected final cutoffs for each bird. Sixteen species showed visible upper-frequency structures that motivated a second filtering branch. The two branches used the same lower cutoff and different upper cutoffs. The wider branch retained the selected upper-frequency structures, and the narrower branch attenuated them. We label these preparation branches ``with harmonics'' and ``without harmonics,'' respectively.

\begin{figure}[htbp]
\centering
\includegraphics[width=.76\linewidth]{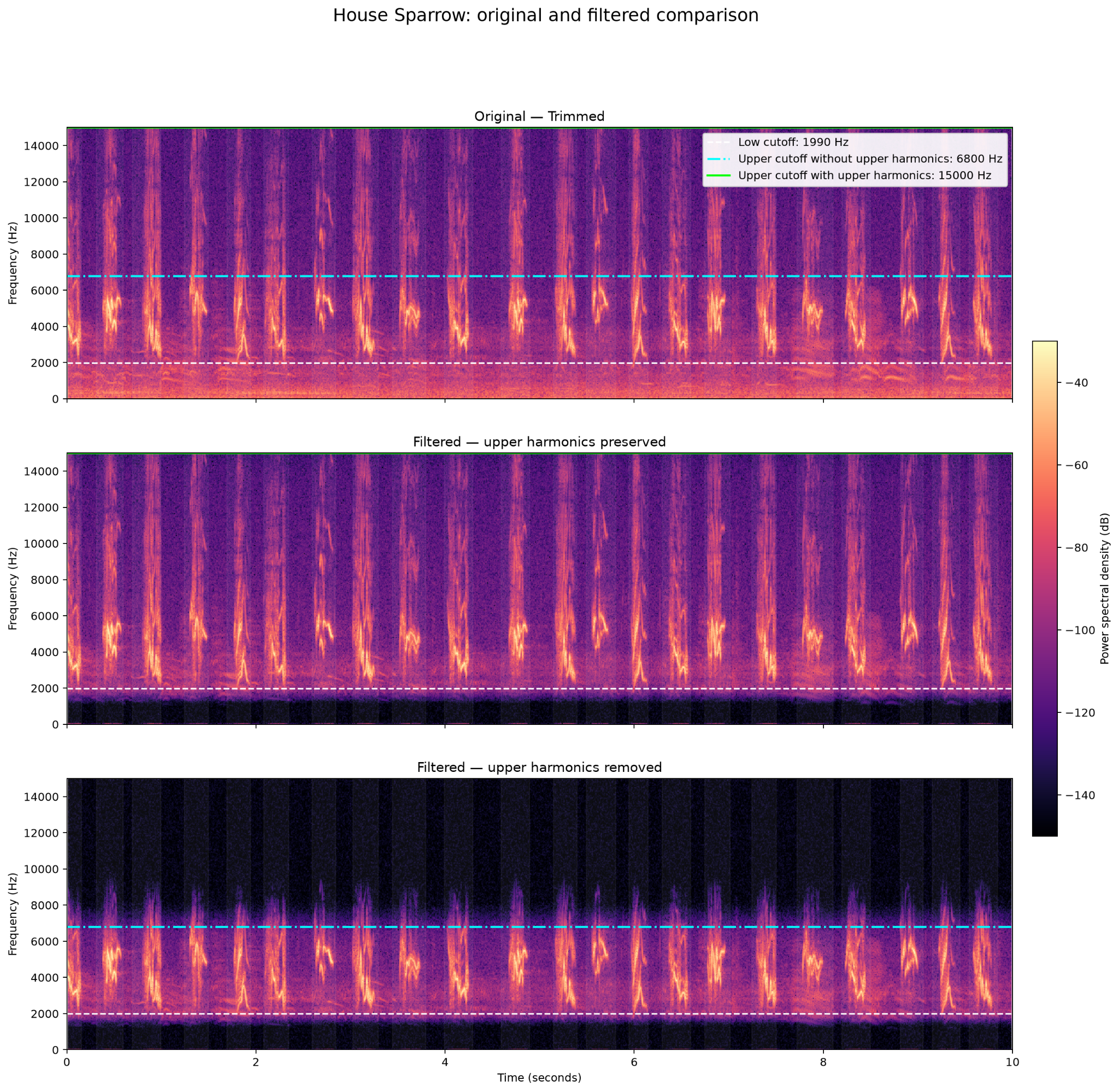}
\caption{House Sparrow stimulus before filtering, after filtering with the upper band retained, and after filtering with the upper band attenuated. Horizontal guides show the selected upper cutoffs.}\label{fig:filtering}
\end{figure}

We applied the configured sixth-order Butterworth filters as second-order sections (SOS) using forward and reverse filtering. This operation preserves the temporal placement of the calls for subsequent use of the annotations. Filtering was applied along the time dimension of the original channel arrangement. Figure~\ref{fig:filtering} illustrates the two branches for House Sparrow. Validation compared power below the lower cutoff, in the common core band, and in the selected upper band. The common core extends from the lower cutoff to the narrower branch's upper cutoff.

\FloatBarrier
\subsection{Channel selection, silencing, and level standardization}
We selected a mono representation from each original trimmed recording. Stereo channels were averaged when their active-call levels differed by at most 3 dB, their correlation was at least 0.85, and the averaged level was within approximately 1 dB of the stronger channel. Otherwise, the stronger active-call channel was selected. The same channel decision was applied to both filtering branches. Existing mono files retained their channel.

\begin{figure}[tb]
\centering
\includegraphics[width=.85\linewidth]{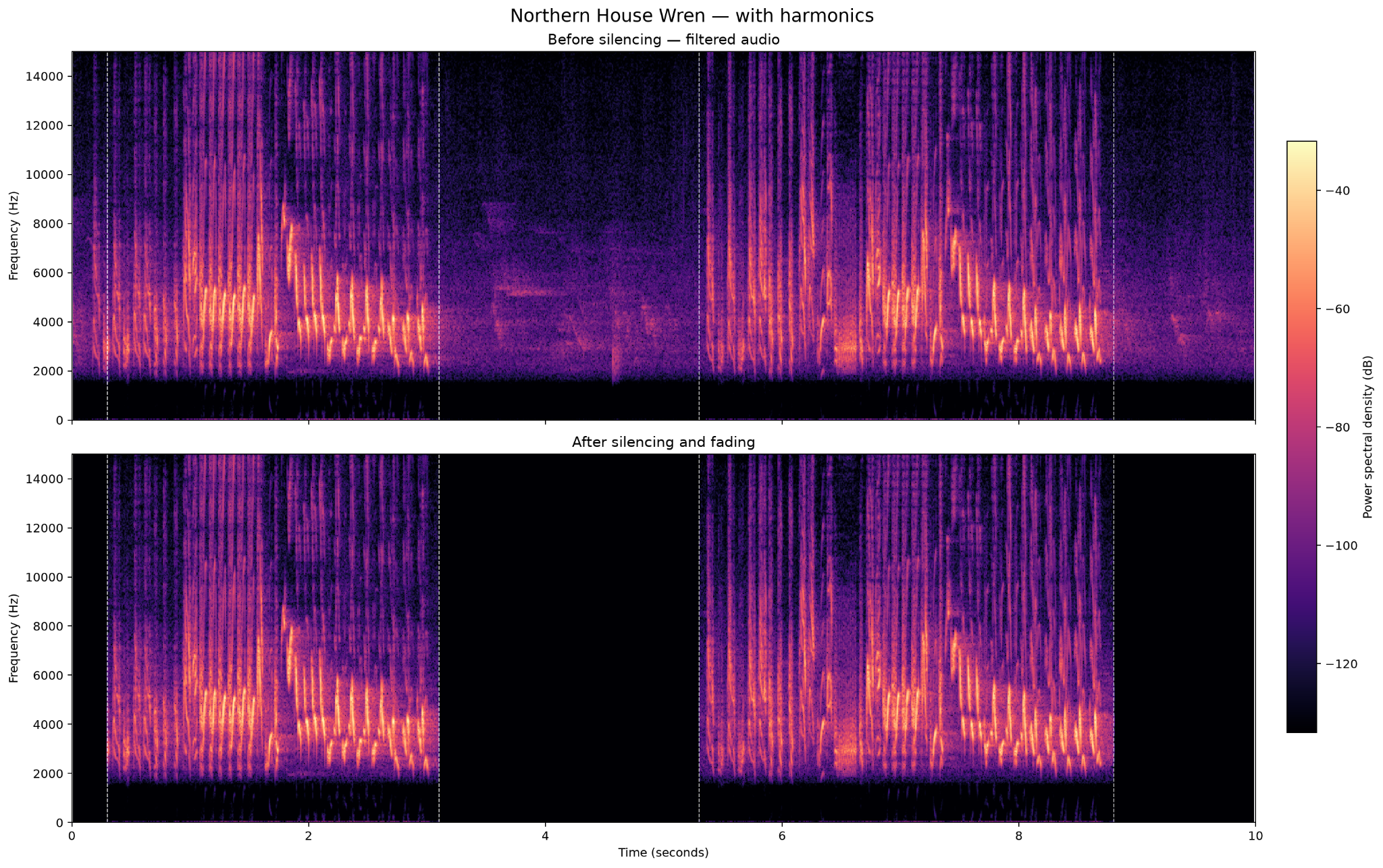}
\caption{Northern House Wren stimulus with the selected upper band retained, before and after silencing and boundary fades. The full ten-second timeline is preserved.}\label{fig:silencing}
\end{figure}

The selected signals were resampled to 44,100 Hz using polyphase resampling and saved as mono PCM16 WAV files. We then set samples outside the active intervals to zero. Linear fades of 8 ms were applied at interval boundaries. For an interval shorter than 16 ms, each fade was limited to half the interval duration. This order places silencing after resampling and preserves digital zeros outside the annotated intervals. Figure~\ref{fig:silencing} shows an example.

The stimulus examples show the effects of the preparation stages. For House Sparrow, the wider filter retains upper-frequency structures that are attenuated in the narrower branch (Figure~\ref{fig:filtering}). For Northern House Wren, silencing removes energy between the annotated call intervals while preserving the temporal placement of the active portions (Figure~\ref{fig:silencing}). The common RMS target sets the digital active-call level for playback. The European Starling peak of approximately $-1.45$ dBFS remains below the specified $-1$ dBFS limit.

\section{Source Recordings}\label{app:sources}
Table~\ref{tab:sources} lists the selected species and Macaulay Library identifiers. These identifiers refer to the source media from which the ten-second stimuli were selected.
\begin{table}[htbp]
\centering
\caption{Source recordings listed in the preparation report.}\label{tab:sources}
\begin{tabular}{ll}
\toprule
Species & Macaulay Library identifier\\
\midrule
American Robin & ML168300\\
Anna's Hummingbird & ML197952\\
Barn Swallow & ML63124\\
Bewick's Wren & ML191106\\
Black Phoebe & ML182034971\\
California Towhee & ML187038\\
Dark-eyed Junco & ML27611071\\
European Starling & ML14861\\
Hooded Oriole & ML245074981\\
House Finch & ML22938\\
House Sparrow & ML138707321\\
Lesser Goldfinch & ML219156941\\
Mourning Dove & ML166991841\\
Northern House Wren & ML107106211\\
Northern Mockingbird & ML534473\\
Northern Yellow Warbler & ML57524601\\
Oak Titmouse & ML88772\\
Rock Pigeon & ML44449551\\
Song Sparrow & ML43973911\\
Spotted Towhee & ML55077\\
\bottomrule
\end{tabular}
\end{table}

\section{Processing and Validation Parameters}\label{app:validation}
Table~\ref{tab:checks} collects the principal parameters of the preparation and acquisition checks. Each processing stage produces measurements tied to the same source species and active intervals. The final per-file statuses identify the outputs satisfying these criteria.
\begin{table}[htbp]
\centering\small
\caption{Configured processing and validation parameters.}\label{tab:checks}
\begin{tabular}{p{.36\linewidth}p{.55\linewidth}}
\toprule
Operation or check & Parameter or criterion\\
\midrule
Welch PSD & 0.10 s Hann window; 50\% overlap\\
Initial spectral range & 100 Hz to source Nyquist frequency\\
Relative PSD cutoff & $-40$ dB relative to peak power\\
Initial energy boundaries & 1st and 99th retained-energy percentiles\\
Suggested boundary margin & 500 Hz at each end, followed by review\\
Filtering & Sixth-order Butterworth design, SOS, forward and reverse\\
Stereo averaging & Level difference $\leq3$ dB, correlation $\geq0.85$, averaging loss approximately $\leq1$ dB\\
Playback output & Mono, 44,100 Hz, PCM16 WAV\\
Resampling duration difference & At most 0.001 s\\
Resampling active RMS change & Within $\pm0.10$ dB\\
Silencing & Zero outside active intervals; 8 ms boundary fades\\
Normalized active-call RMS & $-25$ dBFS target\\
Normalized full-file peak & At most $-1$ dBFS\\
Normalization duration difference & At most $10^{-6}$ s\\
Species recording segment & Mono PCM16, 35,714 Hz, 357,140 frames\\
DC-offset warning & Absolute mean above 7,000 PCM counts\\
Clipping failure & More than 0.1\% at PCM16 limits\\
Near-full-scale failure & More than 1\% at magnitude $\geq32,000$ counts\\
Low-byte diversity warning & Fewer than 32 distinct low-byte values\\
Active BirdNET window & At least 0.25 s overlap with active intervals\\
Heuristic screening criterion & Finite screening score $\geq3$ dB\\
\bottomrule
\end{tabular}
\end{table}

\section{Additional Stimulus Examples}
\begin{figure}[htbp]
\centering
\includegraphics[width=.9\linewidth]{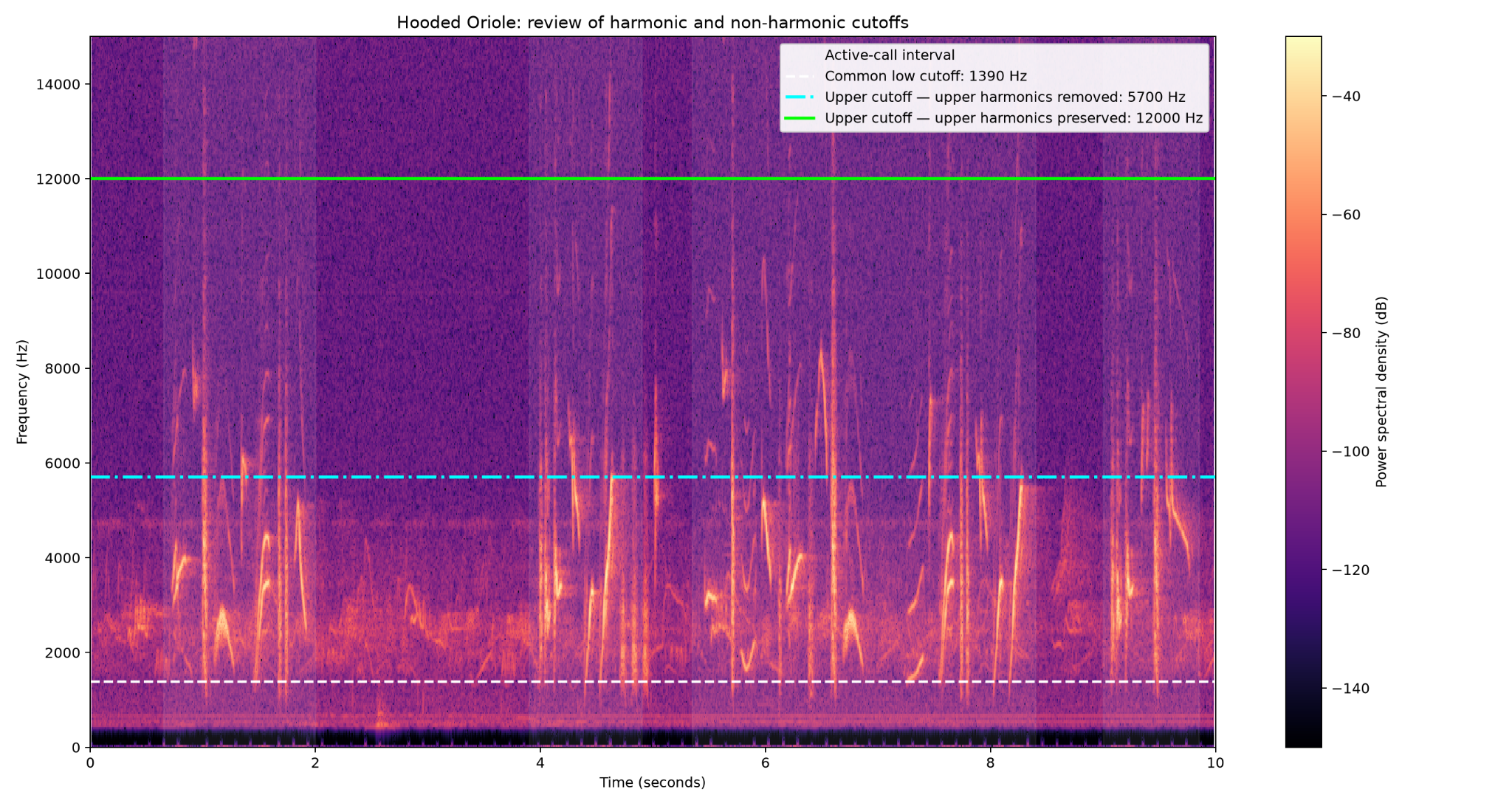}
\caption{Hooded Oriole stimulus with the common lower cutoff and two reviewed upper cutoffs. The horizontal lines locate the selected filtering boundaries relative to the source spectrogram.}\label{fig:cutoffs}
\end{figure}

\begin{figure}[htbp]
\centering
\includegraphics[width=.65\linewidth]{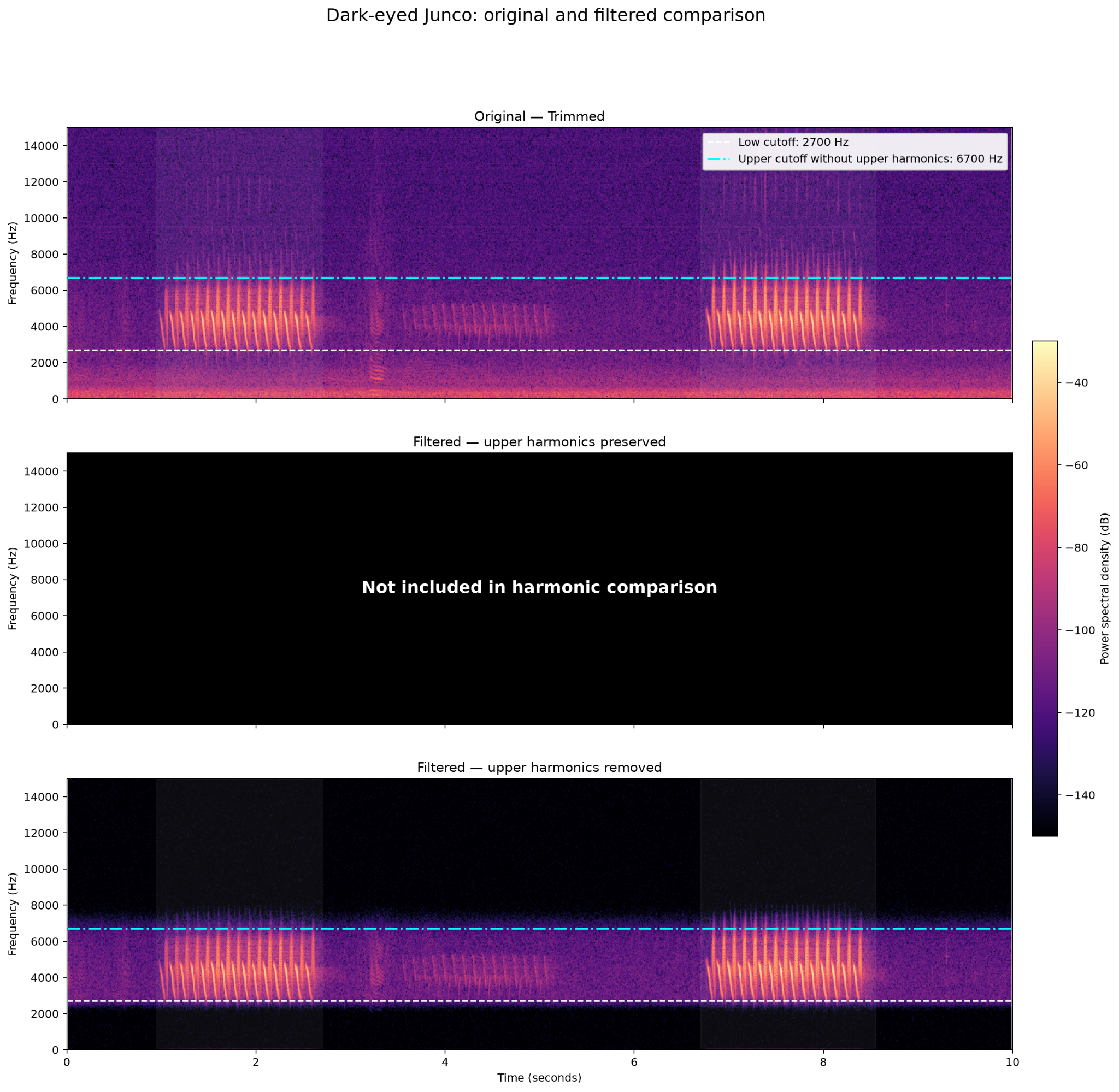}
\caption{Dark-eyed Junco source and filtered stimulus. The middle panel identifies this species as outside the selected harmonic comparison group. The lower panel shows the single filtered branch.}\label{fig:junco}
\end{figure}

Figures~\ref{fig:cutoffs} and~\ref{fig:junco} illustrate selection of upper cutoffs and processing of a species with a single filtering branch. These examples supplement the House Sparrow and Northern House Wren examples in Appendix~\ref{app:preparation}.

\FloatBarrier
\section{BirdNET Confidence by Species and Metadata Configuration}\label{app:confidence-by-configuration}
\begin{figure}[t]
\centering
\includegraphics[width=\linewidth,height=.82\textheight,keepaspectratio]{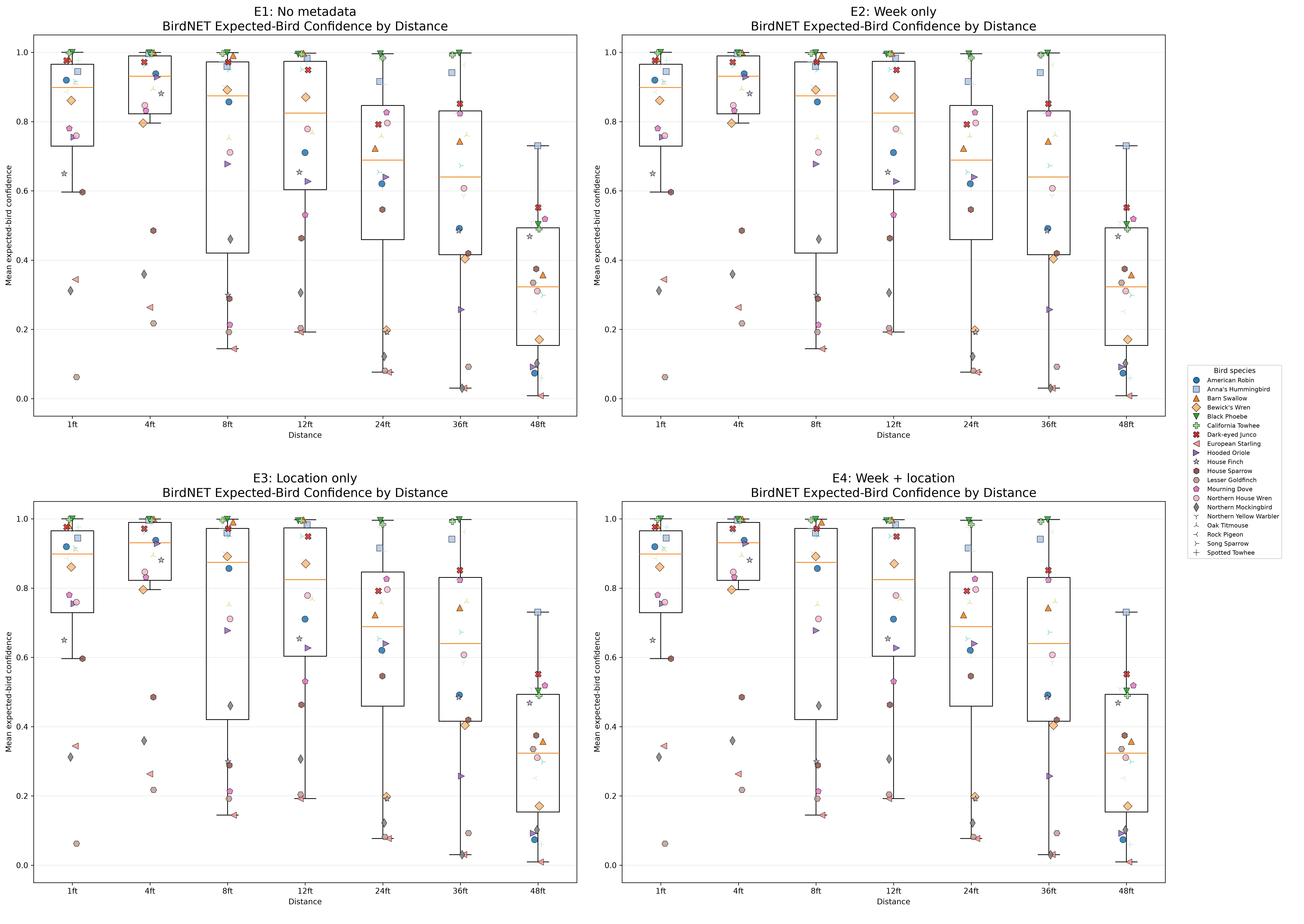}
\caption{Expected-species mean confidence by distance under the four BirdNET metadata configurations: E1, no metadata (top left); E2, week 27 only (top right); E3, Los Angeles location only (bottom left); and E4, week 27 and Los Angeles location (bottom right). Each marker represents one species recording, boxes summarize the 20 species at each distance, and the shared legend applies to all four panels.}\label{fig:confidence-configurations}
\end{figure}

Figure~\ref{fig:confidence-configurations} shows recording-level expected-species mean confidence separately for the four BirdNET metadata configurations. Each panel contains the same 20 species and seven distances. Species retain consistent colors and marker shapes across distances and configurations, allowing species-level confidence changes to be compared without pooling repeated evaluations of the same recordings.

\FloatBarrier

\section{Acoustic Metric Definitions}\label{app:acoustic}
Let $P_{b,d}$ denote the power in band $b$ at distance $d$, and let $d_0=1$ ft. A signed power change relative to the nearest recording can be expressed as
\begin{equation}
 \Delta P_b(d)=10\log_{10}\frac{P_{b,d}}{P_{b,d_0}}.
\end{equation}
A negative value indicates lower received power than at 1 ft. Comparing changes in the high and low bands describes frequency-dependent changes in the received spectrum.

The analysis also calculated the spectral centroid (the power-weighted mean frequency) and the 1st and 99th cumulative-energy percentile frequencies. Their difference gives the bandwidth containing the central 98\% of the measured energy. Changes in these quantities describe movement of the spectral energy and its edges relative to the reference. The Frequency Distortion Index (FDI) is the root-mean-square difference between centered log-power spectra. Let $K$ be the number of retained frequency bins, $S_d[k]$ the PSD at distance $d$ on the 1-ft frequency grid, and $L_d[k]=10\log_{10}\max(S_d[k],10^{-20})$. With $\overline L_d=K^{-1}\sum_{k=1}^{K}L_d[k]$, we calculate
\begin{equation}
 \mathrm{FDI}(d)=\sqrt{\frac{1}{K}\sum_{k=1}^{K}\left[(L_d[k]-\overline L_d)-(L_{d_0}[k]-\overline L_{d_0})\right]^2}.
\end{equation}
Active samples are concatenated before Welch estimation. The Hann window uses the smaller of the active sample count and the nominal 0.10-second window length, with a 256-sample minimum on the nominal length. The overlap is rounded to 50\% of the window, and each window uses constant detrending. PSD estimates use density scaling and a $10^{-20}$ floor. Bins within the selected species frequency range are retained with equal weight for FDI. Target PSDs are linearly interpolated onto the reference grid. Spectral edges use trapezoidal cumulative integration of this restricted PSD, with interpolation at the 1st and 99th percentiles. The attenuation fields store signed target-to-reference power changes, and the high-minus-low field subtracts the low-band change from the high-band change.

We use a heuristic screening score $H_{\mathrm{env}}$ to select a subset for descriptive spectral statistics. Let $P_{\mathrm{active}}$ denote active-interval mean-square amplitude, $P_{\mathrm{rec}}$ the recording-background mean-square estimate, and $P_{\mathrm{env}}$ the source-environment mean-square estimate. Each quantity uses its audio file's digital scale. The calculation is
\begin{align}
 P_{\mathrm{noise}}&=P_{\mathrm{rec}}+P_{\mathrm{env}},\\
 P_{\mathrm{signal}}&=P_{\mathrm{active}}-P_{\mathrm{noise}},\\
 H_{\mathrm{env}}&=10\log_{10}\frac{P_{\mathrm{signal}}}{P_{\mathrm{noise}}}.
\end{align}
The additions and subtraction take place in linear digital power. The score is defined when $P_{\mathrm{noise}}>0$ and $P_{\mathrm{signal}}>0$. We select recordings with finite $H_{\mathrm{env}}\geq3$ dB. The symbols $P_{\mathrm{noise}}$ and $P_{\mathrm{signal}}$ denote the algebraic sum and difference used in this screening calculation.

For each audio file, we remove its full-file mean and estimate recording background as the mean squared samples outside the active intervals and a 0.10-second guard on each side. Active power is the mean squared samples inside the active intervals. The source-environment estimate applies the same procedure to each species' normalized environment-reference audio in the wider filtering branch. Both estimates use full-band time-domain samples at their respective file rates. The source-environment value enters the sum directly and stays fixed across distances. Its scale comes from the normalized source file, while recording-background power uses recorder full-scale units. The source-environment term transfers at unity gain and retains its full-band power. These digital scales and the unity-gain addition define the heuristic screening score. Interpreting the source term as received acoustic noise power would require a calibrated transfer through the playback and recording chain.

\end{document}